\documentclass[pra,twocolumn,aps,floatfix,showpacs,tightenlines,superscriptaddress,amsmath,amssymb,showkeys,10pt,nofootinbib]{revtex4-2}
\usepackage{slashed}
\usepackage{xcolor}
\usepackage{graphicx}
\usepackage{hyperref}

\newcommand{\be}{\begin{equation}}\newcommand{\ee}{\end{equation}}
\newcommand{\bea}{\begin{eqnarray}}\newcommand{\eea}{\end{eqnarray}}
\newcommand{\brr}{\begin{array}}\newcommand{\err}{\end{array}}
\newcommand{\bit}{\begin{itemize}}\newcommand{\eit}{\end{itemize}}
\newcommand{\ben}{\begin{enumerate}}\newcommand{\een}{\end{enumerate}}

\newcommand{\bbm}{\begin{bmatrix}}\newcommand{\ebm}{\end{bmatrix}}
\newcommand{\ba}{\begin{array}}
\newcommand{\ea}{\end{array}}

\newtheorem{mydef}{Definition}
\newtheorem{Lemma}{Lemma}
\newcommand{\bd}{\begin{mydef}} \newcommand{\ed}{\end{mydef}}
\newcommand{\bthe}{\begin{theorem}} \newcommand{\ethe}{\end{theorem}}
\newcommand{\ble}{\begin{Lemma}} \newcommand{\ele}{\end{Lemma}}

\def\lan{\langle}
\def\lf{\left}

\def\non{\nonumber}\def\ran{\rangle}

\def\ri{\right}

\def\De{\Delta}

\def\si{\sigma}

\def\1{{_{1}}}\def\2{{_{2}}}

\def\noHe0{:\;\!\!\;\!\!:H_e(0):\;\!\!\;\!\!:}
\def\noHm0{:\;\!\!\;\!\!:H_\mu(0):\;\!\!\;\!\!:}

\def\lan{\langle}
\def\lf{\left}

\def\non{\nonumber}
\def\ran{\rangle}

\def\ri{\right}

\def\De{\Delta}

\def\si{\sigma}

\def\1{{_{1}}}\def\2{{_{2}}}

\begin{document}

\title{No-signaling-in-time as a condition for macrorealism: the case of neutrino oscillations}

\author{Massimo Blasone}
\email{blasone@sa.infn.it}
\affiliation{Dipartimento di Fisica, Universit\`a di Salerno, Via Giovanni Paolo II 132, 84084 Fisciano (SA), Italy}
\affiliation{INFN Sezione di Napoli, Gruppo collegato di Salerno, Italy}
\author{Fabrizio Illuminati}
\email{filluminati@unisa.it}
\affiliation{INFN Sezione di Napoli, Gruppo collegato di Salerno, Italy}
\affiliation{Dipartimento di Ingegneria Industriale, Universit\`a di Salerno,
Via Giovanni Paolo II 132, 84084 Fisciano (SA), Italy}
\author{Luciano Petruzziello}
\email{lupetruzziello@unisa.it}
\affiliation{INFN Sezione di Napoli, Gruppo collegato di Salerno, Italy}
\affiliation{Dipartimento di Ingegneria Industriale, Universit\`a di Salerno,
Via Giovanni Paolo II 132, 84084 Fisciano (SA), Italy}
\affiliation{Institut f\"ur Theoretische Physik, Albert-Einstein-Allee 11, Universit\"at Ulm, 89069 Ulm, Germany}
\author{Kyrylo Simonov}
\email{kyrylo.simonov@univie.ac.at}
\affiliation{s7 rail technology GmbH, Lastenstra\ss e 36, 4020 Linz, Austria}
%\affiliation{Fakult\"{a}t f\"{u}r Mathematik, Universit\"{a}t Wien, Oskar-Morgenstern-Platz 1, 1090 Vienna, Austria}
\author{Luca Smaldone}
\email{Luca.Smaldone@fuw.edu.pl}
\affiliation{Faculty of Physics, University of Warsaw, ul. Pasteura 5, 02-093 Warsaw, Poland}

\begin{abstract}
We consider two necessary and sufficient conditions for macrorealism recently appeared in the literature, known as no-signaling-in-time and arrow-of-time conditions, respectively, and study them in the context of neutrino flavor transitions, within both the plane wave description and the wave packet approach. We then compare the outcome of the above investigation with the implication of various formulations of Leggett--Garg inequalities. In particular, we show that the fulfillment of the addressed conditions for macrorealism in neutrino oscillations implies the fulfillment of Leggett--Garg inequalities, whereas the converse is not true. Finally, in the framework of wave packet approach, we also prove that, for distances longer than the coherence length, the no-signaling-in-time condition is always violated whilst Leggett--Garg inequalities are not.
\end{abstract}

\maketitle
%%%%%%%%%%%%%%%%%%%%%%%%%%%%%%%%%%%%%%%%%%%%%%%%%%%%
\section{Introduction}
Neutrino mixing and oscillations represent the main indications of physics beyond the Standard Model~\cite{Bilenky:1978nj, Bilenky:1987ty,Beuthe:2001rc,giunti2007fundamentals,Smaldone:2021mii}. Among the multifaceted aspects of the above phenomenon, in recent years the quantum informational properties of mixed flavor states have been widely investigated~\cite{ill1,ill2,ill3,ill4,Dixit:2018gjc,Simonov2019,PhysRevD.100.055021,ill6,Blasone:2021mbc,Simonov2021}. An important achievement along this direction is the characterization of the intrinsic quantum nature of neutrino oscillations, which has been probed with the data available from the MINOS experiment by means of the Leggett--Garg inequalities (LGIs)~\cite{Formaggio2016}.

Loosely speaking, LGIs are typically regarded as the temporal analogues of Bell inequalities; whilst the latter quantify the quantumness of a given system via spatially-separated tests (thus dealing with quantum nonlocality), the former rely on the notion of macroscopic coherence based upon temporal auto-correlation functions~\cite{PhysRevLett.54.857,Brukner2007,Brukner2008,PhysRep2014,Kumari2017}. Indeed, LGIs are closely related to the concept of \emph{macrorealism}, an intuitive view of our classical macroscopic world according to which measurements do not perturb the state of the probed system and reveal a pre-existing, observable quantity. 

Because of their relevance, LGIs have been extensively employed in experimental verifications \cite{Formaggio2016,Wang_2017,Zhang2021,Santini2022,Halliwell2022}. On the same footing, in the last decades systems revealing phenomena of mixing and flavor oscillations have become the subject of an emergent exploration dealing with classicality and macroscopic superpositions~\cite{Bertlmann2003, Donadi2012, Donadi2013, Bahrami2013, SimonovLetter2016, SimonovLetter2016, Simonov2018, Simonov2020, Capolupo2019, gango, Naikoo:2019gme, Naikoo:2020LGI, Dixit2021,Blasone:2021mbc, Wang2022, Simonov2022}. As a matter of fact, it is no coincidence that neutrinos provide a promising probe for testing the validity of LGIs, since their flavor oscillations exhibit quantum coherence even after the particles have traveled macroscopic distances~\cite{gango,Naikoo:2019gme,Naikoo:2020LGI,Dixit2021,Blasone:2021mbc}.

%Another concept, which is related to LGI is \emph{macrorealism}, which tries to capture the main features of classical objects, distinguishing them from the the quantum ones.
%LGI are mainly used which is related to LGI is \emph{macrorealism}, which tries to capture the main features of classical objects, distinguishing them from the the quantum ones.
Despite the pivotal role covered by LGIs, experiments centered around macrorealism reveal a more complex structure if compared with tests based upon local realism~\cite{Clemente2016}. The crucial difference lies in the fact that, whilst Bell inequalities are both necessary and sufficient conditions for local realism \cite{Fine1982}, the fulfillment of LGIs is not in a one-to-one correspondence with macrorealism. Indeed, the validity of the standard LGIs and their variants such as the \emph{Wigner form} of LGIs (WLGIs) \cite{Saha2015} turns out not to be sufficient for macrorealism \cite{Clemente2015,Clemente2016,Halliwel2016}. For this reason, it is essential to introduce another set of conditions for macrorealism which would be both necessary and sufficient; such a set has already been developed, and it is given by a combination of \emph{no-signaling-in-time} (NSIT) (which is an alternative necessary condition for macrorealism~\cite{Brukner2008, Clemente2015}) and \emph{arrow-of-time} (AoT) conditions \cite{Kofler2013,Clemente2015}. Being \textit{equalities} for joint probabilities rather than inequalities, these requirements are more suitable to be interpreted as quantum witnesses. 
%Therefore, in the context of neutrino oscillations, it would be compelling to analyze the interplay between violations of NSIT and violations of LGIs. %, expressed as sets of equalities relating some joint probabilities, are more suitable and can be taken as conditions equivalent to macrorealism.

In this paper, we study the NSIT and AoT conditions in the case of two-flavor neutrino oscillations. We find that, while AoT conditions are always trivially satisfied, neutrino oscillations always violate NSIT excluding an integer set of isolated points. However, if a wave-packet treatment is considered and the measurements are performed at sufficiently large intervals of time (corresponding to distances longer than the coherence length), the NSIT conditions are always violated. This fact confirms that, even after the occurrence of wave-packet decoherence, neutrinos still retain their intrinsic quantum nature, thereby preventing a macrorealistic interpretation of flavor transitions even at late times. In conjunction with that, we also compare the validity of LGIs (WLGIs) with the validity of NSIT and AoT conditions; in so doing, we find that LGIs (WLGIs) are never violated when NSIT and AoT are not, and that for large-time intervals all the LGIs (WLGIs) are fulfilled.
 
The remainder of the paper is organized as follows: in Section~\ref{sec:2}, we review the notion of macrorealism and the related quantifiers we will employ to support our reasoning (namely, LGIs, WLGIs and NSIT/AoT). In Section~\ref{sec:3}, we provide the necessary tools to investigate neutrino oscillations and analyze the ensuing NSIT conditions in the two-flavor approximation; with these results, we then establish a comparison with the predictions stemming from LGIs and WLGIs. Finally, Section \ref{sec:5} contains conclusions and future perspectives.
%%%%%%%%%%%%%%%%%%%%%%%%%%%%%%%%%%%%%%%%%%%%%%
\section{Macrorealism and NSIT conditions}\label{sec:2}

According to our daily experience, we do not observe macroscopic objects around us being in two different positions at the same time. Furthermore, a motionless object with a net vanishing force acting on it stays at all times in a given place which can be determined by simply looking at it. \emph{Macrorealism} aims at formalizing this knowledge by relying on the following basic assumptions\footnote{Often, a third extra condition of \emph{induction} is considered~\cite{PhysRep2014}, which states that the outcome of a measurement on a system cannot be affected by what will/will not be measured on it later.}:
\begin{itemize}
\item
\emph{macrorealism per se}: given a set of available macroscopically distinct states, a macroscopic object is in one of them at any given time;
\item
\emph{non-invasive measurability}: it is possible in principle to determine the state of the macroscopic object without affecting either its state or its dynamical evolution.
\end{itemize}
Similarly to the celebrated Bell inequalities in the framework of local realism, one can derive a set of inequalities (known as LGIs) that have to be satisfied by any physical system abiding by the above macrorealistic prescriptions. To show this in a simple case, let us consider a system with a dichotomous macroscopic observable $O$ with associated values $\pm 1$ which is consecutively measured $N$ times by an observer at fixed time points $\{ t_0, t_1, ..., t_{N-1} \}$. Assuming for simplicity $N=3$ (\emph{i.e.}, three measurements at times $t_0, t_1, t_2$), the measurement statistics with respect to the 2-time correlation functions $C_{ij} = \langle O(t_i) O(t_j) \rangle$ has to satisfy the LGIs~\cite{PhysRep2014,Halliwel2016}
\bea
\label{lgi0121}
  \mathcal{L}_1(t_0, t_1, t_2) = 1+ C_{01} + C_{12} + C_{02} \geq 0 \, , \\[2mm]
\label{lgi0122}
  \mathcal{L}_2(t_0, t_1, t_2) = 1- C_{01} - C_{12} + C_{02} \geq 0 \, , \\[2mm]
  \label{lgi0123}
  \mathcal{L}_3(t_0, t_1, t_2) = 1+ C_{01} - C_{12} - C_{02} \geq 0 \, , \\[2mm]
   \label{lgi0124}
  \mathcal{L}_4(t_0, t_1, t_2) = 1- C_{01} - C_{12} - C_{02} \geq 0 \, ,
\eea

\noindent
if macrorealism holds true. Hence, as for Bell inequalities, these relations can be used to explore the quantumness of a system and the existence of macroscopic superpositions. Indeed, in quantum mechanics the LGIs~(\ref{lgi0121})-\eqref{lgi0124} can be (and are) violated, in particular by the systems coherently oscillating between the states on which $O = \pm 1$, respectively~\cite{PhysRep2014}. 
%at different times $t_0,t_1,t_2$. In quantum theory such measurements are associated with a self-adjoint operator $O(t)$ ($O|\pm\ran = m|\pm\ran$, with $m=\pm 1$). A common popular condition for macrorealism are the LGI \cite{PhysRevLett.54.857}, which are formulated as inequalities involving correlation functions $C_{ij} \equiv \lan O(t_i) O(t_j)\ran$. For example, the inequality

The Leggett--Garg inequalities (\ref{lgi0121})-\eqref{lgi0124} can then be regarded as the temporal counterpart of the Bell inequalities, and just like the latter they are not unique. As a matter of fact, alternative forms of LGIs can be found by focusing solely on the joint probabilities $P(O_i, O_j)$ of finding outcomes $O_i$ and $O_j$ after measuring $O$ at times $t_i$ and $t_j$, respectively (instead of evaluating the functions $C_{ij}$~\cite{Kumari2017,Saha2015}). Indeed, macrorealism entails the existence of an overall joint probability distribution $P(O_0,O_1,O_2)$ of definite outcomes at all measurement times $t_0, t_1, t_2$. Thus, the two-time probabilities $P(O_i, O_j)$ can be straightforwardly calculated as marginals of the overall joint probability distribution. The requirement of positivity $P(O_0,O_1,O_2) \geq 0$ demands specific constraints on $P(O_i, O_j)$; the shape of such constraints can be summarized in the so-called WLGIs \cite{Naikoo:2019gme}
%\footnote{It is easy to check that \eqref{wlgi} is a WLGI. In fact
%
%\bea\nonumber
% P(O_1,O_2)-P(-O_0,O_1)-P(O_0,O_2)  \ = \\[2mm] \non
% -P(-O_0,-O_1,-O_2)-P(O_0,-O_1,O_2) \, , 
%\eea
%
%which should be negative, since the probability distributions $\rho$ are positive. Similar reasonings can be applied to \eqref{wlgi2},\eqref{wlgi3}.} 
\begin{widetext}
\bea \label{wlgi}
\mathcal{W}_1(t_0, t_1, t_2) & = & P(O_1,O_2)-P(-O_0,O_1)-P(O_0,O_2) \ \leq \ 0 \, , \\[2mm]
\mathcal{W}_2(t_0, t_1, t_2) & = & P(O_0,O_2)-P(O_0,-O_1)-P(O_1,O_2) \ \leq \ 0 \, , \label{wlgi2} \\[2mm]
\mathcal{W}_3(t_0, t_1, t_2) & = & P(O_0,O_1)-P(O_1,-O_2)-P(O_0,O_2) \ \leq \ 0 \, . \label{wlgi3}
\eea
\end{widetext}
As it occurs for LGIs (\ref{lgi0121})-\eqref{lgi0124}, WLGIs~(\ref{wlgi})-\eqref{wlgi3} can be violated by quantum mechanical probabilities.

Interestingly, it has been pointed out that all forms of LGIs represent only a necessary (but not a sufficient) condition for macrorealism, which can still be violated even if LGIs are satisfied~\cite{Clemente2015,Clemente2016}. This raises the need to seek alternative conditions that could signal a quantum behavior for the cases in which LGIs provide an incomplete description. A necessary and sufficient condition is given by a set of equalities~\cite{Clemente2015} consisting of two classes that constrain signaling from past to future (known as no-signaling-in-time conditions, or NSIT) and from future to past (known as arrow-of-time conditions, or AoT). In the case $N=3$ (the measurements considered in the present work), one can identify three NSIT conditions
\bea \label{nsit1}
&& \mathrm{NSIT}^{(1)}: \ \ P(O_2) \ = \ \sum_{O_1} P(O_1, O_2) \, , \\[2mm]
&& \mathrm{NSIT}^{(2)}: \ \ P(O_0, O_2) \ = \ \sum_{O_1} P(O_0, O_1, O_2) \, , \label{nsit2} \\[2mm]
&& \mathrm{NSIT}^{(3)}: \ \ P(O_1, O_2) \ = \ \sum_{O_0} P(O_0, O_1, O_2) \, , \label{nsit3}
\eea
and three AoT conditions
\bea
\label{AoT1} && \mathrm{AoT}^{(1)}: \ \ P(O_0, O_1) \ = \ \sum_{O_2} P(O_0, O_1, O_2) \, , \\[2mm] 
\label{AoT2} && \mathrm{AoT}^{(2)}: \ \ P(O_0) \ = \ \sum_{O_1} P(O_0, O_1) \, , \\[2mm]
\label{AoT3} && \mathrm{AoT}^{(3)}: \ \ P(O_1) \ = \ \sum_{O_2} P(O_1, O_2) \, .
\eea
Remarkably, it can be proved that NSIT conditions imply all possible forms of LGIs. 

In the following, we apply the notions introduced above in the context of neutrino flavor transitions to compare the different conditions for macrorealism. %in the complete QFT framework previously introduced. Moreover, we will compare the complete QFT results with the approximate QM ones, which are only valid in the relativistic limit.
%%%%%%%%%%%%%%%%%%%%%%%%%%%%%%%%%%%%%%%%%%%%%%%%%%%%%%%%%%%%%%%%%%%%%%%
\section{Macrorealism in neutrino oscillations} \label{sec:3}
%%%%%%%%%%%%%%%%%%%%%%%%%%%%%%%%%%%%%%%%%%%%%%%%%%%%%%%%%%%%%%
\subsection{Phenomenology of neutrino oscillations} 
Neutrinos provide a paradigmatic example of mixed particles, whose physical (flavor) states distinguishable in a weak process do not coincide with the (mass) eigenstates of their Hamiltonian, which propagate with frequencies that depend on the corresponding masses. In the relativistic regime, neutrino mass eigenstates evolve according to 
\begin{eqnarray}
|\nu_j(t)\ran \ &=& \ e^{-i E_j t} |\nu_j(0)\ran, \\
E_j &=& \sqrt{p^2+m_j^2} \; \approx \; E + \frac{m_j}{2E},
\end{eqnarray}
where the masses $m_j$ are taken to be much smaller than their momentum and $E = p$ is the energy of a massless neutrino. On the other hand, flavor states are well-described as superpositions of the mass eigenstates~\cite{Bilenky:1978nj,Bilenky:1987ty}
\be
|\nu_\si(t) \ran \ = \ \sum_{j} \, U^*_{\si j} |\nu_j(t)\ran \, , 
\ee
with coefficients given by the elements $U_{\si j}$ of the mixing matrix $U$. The non-equivalence of physical flavor states and mass eigenstates of the particle Hamiltonian ascribed to the mixing phenomenon is responsible for the oscillation between distinct flavor states. If a neutrino is produced in a weak process at time $t = 0$ with a given flavor $\sigma$, it evolves into a superposition of flavor states at $t>0$ in such a way that the probability of detecting another flavor $\rho$ is
%From the above relations
%%
%\be
%|\nu_\si(t) \ran \ = \ \sum_\rho \, \lf(\sum_{j} \,U_{\rho j} U^*_{\si j} e^{-i E_j t}\ri) |\nu_\rho(0)\ran \, , 
%\ee
%%
%The flavor-oscillation probability is thus
%
\bea \non
P_{\si \rightarrow \rho}(t) & = & \lf|\lan \nu_\rho(t)|\nu_\si(0)\ri|^2 \\
 & = & \sum_{j,k} \, U_{\rho j} U_{\si k} U^*_{\rho k} U^*_{\si j} \exp\lf(-i \frac{\Delta m^2_{j k}}{2 E} t\ri) \, , 
\eea
where $\Delta m^2_{j k} \equiv m_j^2-m_k^2$. In particular, for the two-flavor case (a typical approximation that successfully describes many experiments with good accuracy \cite{Bilenky2008}), the mixing matrix is given by
\be
U \ = \ \begin{pmatrix}
\cos \theta & \sin \theta \\
-\sin \theta & \cos \theta 
\end{pmatrix} \, , 
\ee
with $\theta$ being the \emph{mixing angle}. Under these assumptions, the flavor oscillation probability is given by the Pontecorvo formula
\bea
\label{OscPW} P_{\si \rightarrow \rho}(t) & = & \sin^{2}(2 \theta) \, \sin^2 \lf(\frac{\De m^2}{2 E} t\ri) \, , \quad \si \neq \rho \, , \\[2mm] \label{eqosc}
P_{\si \rightarrow \si}(t) & = & 1-P_{\si \rightarrow \rho}(t) \, , 
\eea
and $\De m^2 \equiv \De m^2_{12}$. In light of these features, flavor neutrinos resemble the behavior of two-level systems such as  spin-${1}/{2}$ states and polarized photons; hence, they are naturally liable to be studied in the framework of macrorealism.

Note that, in the scenario described so far, mass eigenstates possess a definite momentum $p$; therefore, they are considered as propagating plane waves. Nevertheless, the above picture still manages to fit most of neutrino physics phenomenology that is probed in actual experiments. However, a more realistic investigation of neutrinos requires a treatment of mass eigenstates in terms of wave packets. To this aim, let us now consider a neutrino propagating along the $x$-direction
\be
|\nu_\si(x,t)\ran \ = \ \sum_j \, U^*_{\si j} \, \psi_j(t,x) \, |\nu_j\ran \, ,
\ee
where the wave packets $\psi_j(t,x)$ can be chosen as being Gaussian functions~\cite{Giunti1991}
\bea \non
\psi_j(t,x) & = & \lf(\sqrt{2 \pi} \si_x \ri)^{-\frac{1}{2}} e^{i (p x -E_j  t)} e^{-\frac{(x-v_j t)^2}{4 \si^2_x}} \, .
\eea
Here, $p$ is the average momentum of the wave packet\footnote{To keep our considerations simple and without loss of generality, we can impose the same average momentum for all mass eigenstates $|\nu_j\rangle$.}, while $\si_x$ is the spatial spreading and
\begin{eqnarray}
v_j = \frac{p}{E_j} \; \approx \; 1 - \frac{m_j^2}{2 E^2} \, , 
\end{eqnarray}
where $v_j$ are the group velocities. The flavor oscillation formula is thus given by
\bea \non
P_{\si \to \rho}(t,x) 
 & = & \lf(\sqrt{2 \pi} \si_x \ri)^{-1} \, \sum_{j,k} \, U_{\rho j} U_{\si k} U^*_{\rho k} U^*_{\si j} e^{-i \frac{\Delta m^2_{j k}}{2 E} t}\\
&\times & e^{-\frac{(x-v_j t)^2}{4 \si^2_x}-\frac{(x-v_k t)^2}{4 \si^2_x}} \, . \label{oscdam}
\eea
In neutrino experiments, there is no direct access to time measurements, but the distance between the source and the detector is known. Therefore, concerning neutrino phenomenological studies, time is typically superseded by space. As our aim is to test macrorealism involving measurements taken at different times, a reverse conversion of space into time is mandatory. This procedure does not affect the oscillation formula, which essentially remains the same because of the interchangeability between time and space in the relativistic regime \cite{Giunti1991}. 

Now, we can integrate (\ref{oscdam}) over $x$ and normalize it in order to obtain a consistent probabilistic description (\emph{i.e.}, $\sum_\si P_{\si \to \rho}(t)=1$). Eventually, one obtains the following oscillation formula:
\bea \non
P_{\si \rightarrow \rho}(t) 
 & = & \sum_{j,k} \, U_{\rho j} U_{\si k} U^*_{\rho k} U^*_{\si j} \exp\lf(-i \frac{\Delta m^2_{j k}}{2 E} t\ri)\\
 & \times & \exp\lf(-\frac{ (\De m^2_{j k})^2 \, t^2}{32 E^4 \si^2_x}\ri) \, . \label{oscdam1}
\eea
The exponential damping factor is responsible for the relative spread of mass-neutrino wave packets and, in turn, for the decoherence mechanism which averages the oscillations on long-time intervals (distances). Therefore, it is possible to identify a characteristic space/time scale at which the decoherence occurs, namely the so-called \emph{coherence length}
\be
L^{coh}_{jk} \ = \ \frac{4\sqrt{2}\,E^2}{\lf|\Delta m_{jk}^2\ri|}\sigma_x \, .
\ee
Finally, by specializing Eq. \eqref{oscdam1}  for the two-flavor case, the oscillation formula reads
\be
P_{\si \rightarrow \rho}(t) \ = \ \frac{\sin^{2}(2 \theta)}{2} \, \lf(1- e^{-\lf(\frac{t}{L^{coh}}\ri)^2}\cos \lf(\frac{\De m^2}{E} t\ri)\ri) \, , \label{OscWP}
\ee
with $L^{coh} = \frac{4\sqrt{2}\,E^2}{\lf|\Delta m^2\ri|}\sigma_x$. 

We are now ready to introduce the necessary and sufficient conditions for macrorealism in neutrino oscillations. For this purpose, both the plane-wave and the wave-packet description of two-flavor Dirac neutrinos will be considered.

%%%%%%%%%%%%%%%%%%%%%%%%%%%%%%%%%%%%%%%%%%%%%%%%%%%%%%%%%%%%%%%%%%%%%%%%%%%%%%
\subsection{Necessary and sufficient NSIT/AoT conditions for macrorealism in neutrino oscillations}

\begin{figure}
\centering
\hspace{-5mm}    \includegraphics[width=9cm]{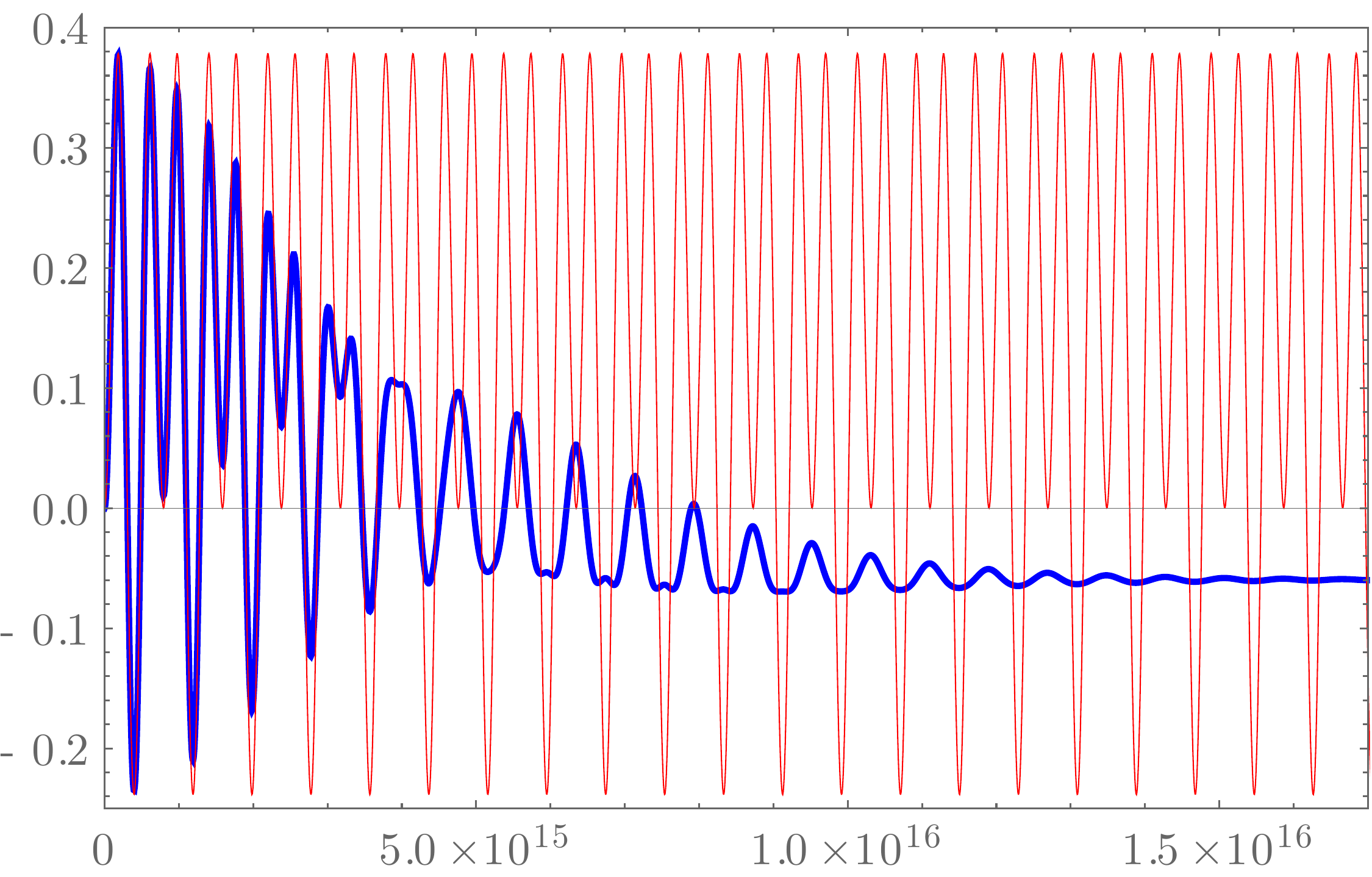}
    \caption{$\mathcal{N}(t)$ for the plane-wave (red) and the Gaussian wave-packet (blue) approach as a function of time expressed in eV$^{-1}$. The values used to generate the plot have been taken from the MINOS experiment \cite{data}, with $\sin^2\theta=0.314$, $\Delta m^2=7.92\times10^{-5}$ eV$^2$, $E=10$ GeV and $\sigma_x=0.5$ GeV$^{-1}$.}
    \label{fig:my_nsitplwave}
\end{figure}

In order to test macrorealism in neutrino oscillations using the combined NSIT/AoT conditions~(\ref{nsit1})--(\ref{AoT3}), we choose neutrino flavor to be the macroscopic dichotomous observable $O(t)$. Since we work within the two-flavor approximation (where the flavor can be either electronic $e$ or muonic $\mu$), we define it as $O(t) = |\nu_e(t) \ran \lan \nu_e(t)|-|\nu_\mu(t) \ran \lan \nu_\mu(t)|$, which thus represents a dichotomous variable with values $\pm 1$ corresponding to $e$- and $\mu$-neutrino flavors, respectively. The ensuing joint probabilities in the NSIT/AoT conditions~(\ref{nsit1})--(\ref{AoT3}) for the measurement outcomes can be straightforwardly rewritten in terms of flavor oscillating probabilities using the conditional probability rule
\begin{eqnarray}
 \nonumber P(O_i, O_j) &=& P(O_i) P(O_j | O_i) \\
 &=& P_{O_0 \rightarrow O_i}(t_i) P_{O_i \rightarrow O_j}(t_j - t_i).
\end{eqnarray}
Without loss of generality, we assume that an electronic neutrino is produced at time $t_0=0$ and its flavor is subsequently measured at $t_1=t$ and $t_2 = 2t$. When the measurement outcomes $O_i$ are fixed, we assume $O_0 = +1 \equiv e$, $O_1 = -1 \equiv \mu$, and $O_2 = -1 \equiv \mu$. Therefore, the full set of NSIT/AoT conditions in neutrino oscillations is
\begin{widetext}
\begin{equation*}
\begin{aligned}
P_{e\rightarrow \mu}(2t) &= P_{e\rightarrow e}(t)P_{e\rightarrow \mu}(t) + P_{e\rightarrow \mu}(t)P_{\mu\rightarrow \mu}(t)\\
P_{e\rightarrow e}(0) P_{e\rightarrow \mu}(2t) &= P_{e\rightarrow e}(0) P_{e\rightarrow e}(t)P_{e\rightarrow \mu}(t) + P_{e\rightarrow e}(0) P_{e\rightarrow \mu}(t)P_{\mu\rightarrow \mu}(t)\\
P_{e\rightarrow \mu}(t)P_{\mu\rightarrow \mu}(t) &= P_{e\rightarrow e}(0) P_{e\rightarrow \mu}(t)P_{\mu\rightarrow \mu}(t) + P_{e\rightarrow \mu}(0) P_{e\rightarrow \mu}(t)P_{\mu\rightarrow \mu}(t) \\
\nonumber P_{e \rightarrow e}(0) P_{e\rightarrow \mu}(t) &= P_{e \rightarrow e}(0) P_{e\rightarrow \mu}(t)P_{\mu\rightarrow e}(t) + P_{e \rightarrow e}(0) P_{e\rightarrow \mu}(t)P_{\mu\rightarrow \mu}(t) \\
\nonumber P_{e \rightarrow e}(0) &= P_{e\rightarrow e}(0) P_{e\rightarrow e}(t) + P_{e\rightarrow e}(0) P_{e\rightarrow \mu}(t) \\
\nonumber P_{e\rightarrow \mu}(t) &= P_{e\rightarrow \mu}(t)P_{\mu\rightarrow e}(t) + P_{e\rightarrow \mu}(t)P_{\mu\rightarrow \mu}(t) 
\end{aligned}
\begin{aligned}
&\left.\vphantom{\begin{aligned}
P_{e\rightarrow \mu}(2t) &= P_{e\rightarrow e}(t)P_{e\rightarrow \mu}(t) + P_{e\rightarrow \mu}(t)P_{\mu\rightarrow \mu}(t)\\
P_{e\rightarrow e}(0) P_{e\rightarrow \mu}(2t) &= P_{e\rightarrow e}(0) P_{e\rightarrow e}(t)P_{e\rightarrow \mu}(t) + P_{e\rightarrow e}(0) P_{e\rightarrow \mu}(t)P_{\mu\rightarrow \mu}(t)\\
P_{e\rightarrow \mu}(t)P_{\mu\rightarrow \mu}(t) &= P_{e\rightarrow e}(0) P_{e\rightarrow \mu}(t)P_{\mu\rightarrow \mu}(t) + P_{e\rightarrow \mu}(0) P_{e\rightarrow \mu}(t)P_{\mu\rightarrow \mu}(t)
\end{aligned}}\right\rbrace\quad\text{NSIT}\\
&\left.\vphantom{\begin{aligned}
\nonumber P_{e \rightarrow e}(0) P_{e\rightarrow \mu}(t) &= P_{e \rightarrow e}(0) P_{e\rightarrow \mu}(t)P_{\mu\rightarrow e}(t) + P_{e \rightarrow e}(0) P_{e\rightarrow \mu}(t)P_{\mu\rightarrow \mu}(t) \\
\nonumber P_{e \rightarrow e}(0) &= P_{e\rightarrow e}(0) P_{e\rightarrow e}(t) + P_{e\rightarrow e}(0) P_{e\rightarrow \mu}(t) \\
\nonumber P_{e\rightarrow \mu}(t) &= P_{e\rightarrow \mu}(t)P_{\mu\rightarrow e}(t) + P_{e\rightarrow \mu}(t)P_{\mu\rightarrow \mu}(t)
\end{aligned}}\right\rbrace\quad\text{AoT}
\end{aligned}
\end{equation*}
\end{widetext}
Interestingly, by suitably manipulating the AoT conditions, one ends up with three relations which are identically satisfied at all times\footnote{Note that this occurrence might not be true when considering the three-flavor scenario because of the presence of a non-vanishing CP-violating phase.}, that is 
\bea
\label{AoT1:QM} && \mathrm{AoT}^{(1)}: \ \ 1 \ = \ P_{\mu\rightarrow e}(t) + P_{\mu\rightarrow \mu}(t) \, , \\[2mm] 
\label{AoT2:QM} && \mathrm{AoT}^{(2)}: \ \ 1 \ = \ P_{e\rightarrow e}(t) + P_{e\rightarrow \mu}(t) \, , \\[2mm]
\label{AoT3:QM} && \mathrm{AoT}^{(3)}: \ \ 1 \ = \ P_{\mu\rightarrow e}(t) + P_{\mu\rightarrow \mu}(t) \, .
\eea
%AoT$^{(1)}$:
%\begin{eqnarray}
%\nonumber P_{01}(e, \mu) &=& P_{012}(e, \mu, e) + P_{012}(e, \mu, \mu)\\
%\nonumber P_{e\rightarrow \mu}(t) &=& P_{e\rightarrow \mu}(t)P_{\mu\rightarrow e}(t) + P_{e\rightarrow \mu}(t)P_{\mu\rightarrow \mu}(t) \\
%\label{AoT1:QM} 1 &=& P_{\mu\rightarrow e}(t) + P_{\mu\rightarrow \mu}(t).
%\end{eqnarray}
%
%AoT$^{(2)}$:
%\begin{eqnarray}
%\nonumber P_0(e) &=& P_{01}(e, e) + P_{01}(e, \mu)\\
%\label{AoT2:QM} 1 &=& P_{e\rightarrow e}(t) + P_{e\rightarrow \mu}(t) .
%\end{eqnarray}
%
%AoT$^{(3)}$:
%\begin{eqnarray}
%\nonumber P_1(\mu) &=& P_{12}(\mu, e) + P_{12}(\mu, \mu)\\
%\nonumber P_{e\rightarrow \mu}(t) &=& P_{e\rightarrow \mu}(t)P_{\mu\rightarrow e}(t) + P_{e\rightarrow \mu}(t)P_{\mu\rightarrow \mu}(t) \\
%\label{AoT3:QM} 1 &=& P_{\mu\rightarrow e}(t) + P_{\mu\rightarrow \mu}(t).
%\end{eqnarray}
%
This is somewhat expected, because the AoT conditions are usually satisfied in standard quantum mechanics~\cite{Clemente2016}. Thus, for neutrino oscillations, AoT conditions can be safely neglected, thereby leaving the NSIT conditions as the relevant ones. Accounting for the symmetry of flavor oscillation probabilities under exchange of flavors, \emph{i.e.}, $P_{e\rightarrow e}(t) = P_{\mu\rightarrow \mu}(t)$ and $P_{e\rightarrow \mu}(t) = P_{\mu\rightarrow e}(t)$, the NSIT are then given by
\bea
\label{NSIT1:QM} && \mathrm{NSIT}^{(1)}: \ \ P_{e\rightarrow \mu}(2t) \ = \ 2P_{e\rightarrow \mu}(t)P_{e\rightarrow e}(t) \, , \\[2mm] 
\label{NSIT2:QM} && \mathrm{NSIT}^{(2)}: \ \ P_{e\rightarrow \mu}(2t) \ = \ 2P_{e\rightarrow \mu}(t)P_{e\rightarrow e}(t) \, , \\[2mm]
\label{NSIT3:QM} && \mathrm{NSIT}^{(3)}: \ \ P_{e\rightarrow \mu}(t)P_{\mu\rightarrow \mu}(t) \ = \ P_{e\rightarrow \mu}(t)P_{\mu\rightarrow \mu}(t).
\eea
%
%NSIT$^{(1)}$:
%\begin{eqnarray}
%P_{2}(\mu) &=& P_{12}(e, \mu) + P_{12}(\mu, \mu)\\
%P_{e\rightarrow \mu}(2t) &=& P_{e\rightarrow e}(t)P_{e\rightarrow \mu}(t) + P_{e\rightarrow \mu}(t)P_{\mu\rightarrow \mu}(t)\\
%P_{e\rightarrow \mu}(2t) &=& 2P_{e\rightarrow \mu}(t)P_{e\rightarrow e}(t) 
%\end{eqnarray}
%
%NSIT$^{(2)}$:
%\begin{eqnarray}
%P_{02}(e, \mu) &=& P_{012}(e, e, \mu) + P_{012}(e, \mu, \mu)\\
%P_{e\rightarrow \mu}(2t) &=& P_{e\rightarrow e}(t)P_{e\rightarrow \mu}(t) + P_{e\rightarrow \mu}(t)P_{\mu\rightarrow \mu}(t)\\
%P_{e\rightarrow \mu}(2t) &=& 2P_{e\rightarrow \mu}(t)P_{e\rightarrow e}(t) 
%\end{eqnarray}
%
%NSIT$^{(3)}$:
%\begin{eqnarray}
%P_{12}(\mu, \mu) &=& P_{012}(e, \mu, \mu) + P_{012}(\mu, \mu, \mu)\\
%P_{e\rightarrow \mu}(t)P_{\mu\rightarrow \mu}(t) &=& P_{e\rightarrow \mu}(t)P_{\mu\rightarrow \mu}(t) 
%\end{eqnarray}
%
It is straightforward to check that $\mathrm{NSIT}^{(3)}$ is a trivial relation, whilst $\mathrm{NSIT}^{(1)}$ and $\mathrm{NSIT}^{(2)}$ coincide. Consequently, macrorealism in neutrino oscillations can be witnessed by a single necessary and sufficient NSIT condition:
\begin{equation}
\mathcal{N}(t) \ \equiv \  P_{e\rightarrow\mu}(2t) \,  - \,  2P_{e \rightarrow\mu}(t) \, P_{e\rightarrow e}(t) \ = \ 0. \label{NSIT:neutrino}
\end{equation}
%which draw together the sufficient and necessary conditions for the macrorealism. 
The function $\mathcal{N}$ is plotted in Fig.~\ref{fig:my_nsitplwave} as a function of time for both the plane-wave and the Gaussian wave-packet flavor oscillation probabilities~\eqref{OscPW} and~\eqref{OscWP}, respectively. It is worth stressing that, for the plane-wave description, the NSIT condition~\eqref{NSIT:neutrino} is periodically fulfilled in isolated points. On the other hand, in the realistic wave-packet scenario, Eq. \eqref{NSIT:neutrino} is fulfilled only for fewer values of the time with respect to the previous case; this occurs because the behavior of $\mathcal{N}(t)$ is similar to the plane-wave result only for small $t$ (\emph{i.e.}, when the damping exponential is still close to unity). Nevertheless, it is crucial to observe that, for large $t$, $\mathcal{N}(t)$ approaches a constant value which in general is different from zero, thereby preventing flavor transitions to be interpreted in a macrorealistic way. This occurs because, at late times, one can check that
\begin{equation}\label{limit}
\lim_{t\to+\infty} \mathcal{N}(t)=-\frac{\sin^2(4\theta)}{8}\,,   
\end{equation}
which is identically zero only for integer multiples of the maximal mixing angle $\pi/4$.

The above picture can be easily explained in quantum informational terms. Indeed, if neutrinos with different flavors are viewed as being qubits of a two-level system \cite{ill1,ill2,ill3,ill4}, it can be shown that, despite the decoherence due to the wave-packet spreading, the amount of quantum correlations shared by the qubits is always non-vanishing, thus allowing for the constant presence of a signature of quantum behavior \cite{correlation,correlation2}. In turn, this fact entails that, regardless of the distance traveled and of the wave-packet separation, for realistic values of the mixing angle (such as the ones used in Fig. \ref{fig:my_nsitplwave} \cite{data}) under no circumstances the phenomenon of flavor transition is compatible with macrorealism.

Before concluding this section, an important remark has to be made: the obtained results related to the NSIT/AoT conditions for macrorealism in neutrino oscillations are independent of the choice of the initial condition (namely, the neutrino flavor at $t=0$) and the values of the outcomes $O_0$, $O_1$, and $O_2$. In fact, by following the same steps as above, one can easily prove that any arbitrary choice for $O_0$, $O_1$, and $O_2$ leads to the same necessary and sufficient condition~\eqref{NSIT:neutrino}. This statement further corroborates the reliability of neutrino oscillations as a suitable instrument to investigate macrorealism.

%%%%%%%%%%%%%%%%%%%%%%%%%%%%%%%%%%%%%%%%%%%%%%%%%%%%%%%%%%%%
\subsection{Comparison of NSIT/AoT with other conditions for macrorealism}
\begin{figure}
\centering
\hspace{-5mm}    
    \includegraphics[width=9cm]{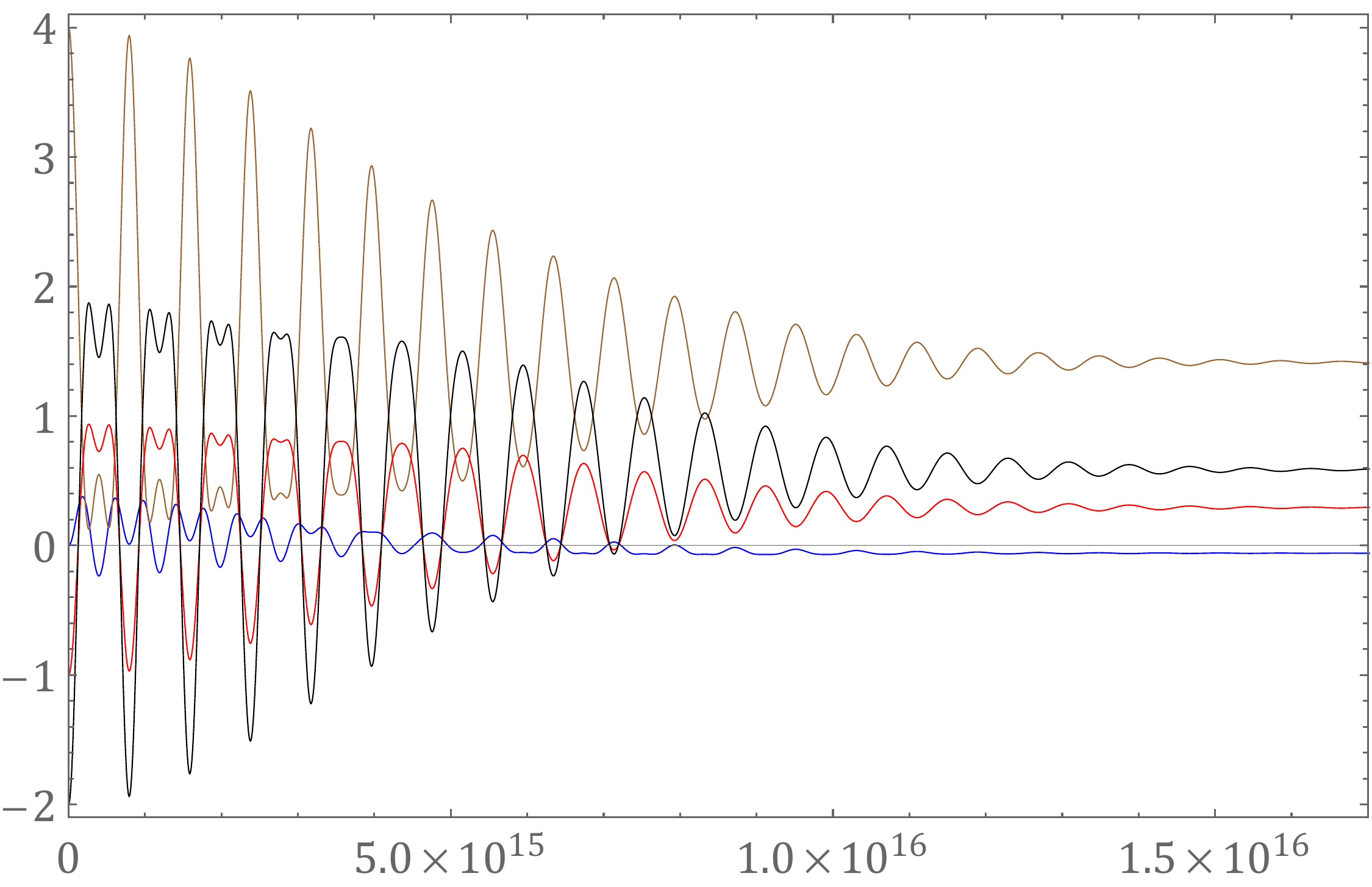}
    \caption{$\mathcal{N}(t)$ (blue) vs $\mathcal{L}_1(t)$ (brown), $\mathcal{L}_2(t)$ (red), and $\mathcal{L}_4(t)$ (black) as functions of time expressed in eV$^{-1}$. The former witnesses violation of the NSIT condition whenever it is not equal to zero, while the latter witness violation of LGIs whenever they are negative. It is worth highlighting that, for large $t$, all $\mathcal{L}_j(t)$ are always non-negative while $\mathcal{N}(t)$ differs from zero. The values used to generate the plot have been taken from the MINOS experiment \cite{data}, with $\sin^2\theta=0.314$, $\Delta m^2=7.92\times10^{-5}$ eV$^2$, $E=10$ GeV and $\sigma_x=0.5$ GeV$^{-1}$.}
    \label{fig:my_nsitvslgi}
\end{figure}

In order to compare the condition (\ref{NSIT:neutrino}) of macrorealism in neutrino oscillations with the predictions obtained with LGIs, we have to adapt the latter to the problem at hand. To this aim, we first investigate the LGIs in their standard formulations \eqref{lgi0121}-\eqref{lgi0124}, which require the evaluation of the correlation functions in terms of flavor oscillation probabilities:
\begin{eqnarray}
    \nonumber C_{ij} &=& \langle O(t_i) O(t_j) \rangle \\
    \nonumber &=& P_{e \rightarrow e}(t_i) \Bigl( P_{e \rightarrow e}(t_j-t_i) -  P_{e \rightarrow \mu}(t_j-t_i) \Bigr) \\
    &+& P_{e \rightarrow \mu}(t_i) \Bigr( P_{\mu \rightarrow \mu}(t_j-t_i) - P_{\mu \rightarrow e}(t_j-t_i) \Bigr).
\end{eqnarray}
Bearing this in mind, we have
\begin{eqnarray}
    \nonumber C_{01} &=& P_{e \rightarrow e}(0) \Bigl( P_{e \rightarrow e}(t) - P_{e \rightarrow \mu}(t)\Bigr) \\
    \nonumber &+& P_{e \rightarrow \mu}(0) \Bigl( P_{\mu \rightarrow \mu}(t) - P_{\mu \rightarrow e}(t)\Bigr), \\
    \nonumber C_{12} &=& P_{e \rightarrow e}(t) \Bigl( P_{e \rightarrow e}(t) - P_{e \rightarrow \mu}(t)\Bigr) \\
    \nonumber &+& P_{e \rightarrow \mu}(t) \Bigl( P_{\mu \rightarrow \mu}(t) - P_{\mu \rightarrow e}(t)\Bigr), \\
    \nonumber C_{02} &=& P_{e \rightarrow e}(0) \Bigl( P_{e \rightarrow e}(2t) - P_{e \rightarrow \mu}(2t)\Bigr) \\
    \nonumber &+& P_{e \rightarrow \mu}(0) \Bigl( P_{\mu \rightarrow \mu}(2t) - P_{\mu \rightarrow e}(2t)\Bigr).
\end{eqnarray}
Finally, invoking the symmetry of flavor oscillation probabilities under the exchange of flavor subscripts, it is immediate to verify that
\begin{eqnarray}
    \label{C01} C_{01} &=& P_{e \rightarrow e}(t) - P_{e \rightarrow \mu}(t), \\
    \label{C12} C_{12} &=& P_{e \rightarrow e}(t) - P_{e \rightarrow \mu}(t), \\
    \label{C02} C_{02} &=& P_{e \rightarrow e}(2t) - P_{e \rightarrow \mu}(2t).
\end{eqnarray}
Now, plugging the found correlation functions (\ref{C01})--(\ref{C02}) into the definitions (\ref{lgi0121})-\eqref{lgi0124}, we reach the expression of LGIs in the framework of neutrino oscillations, that is
%\begin{equation}
%
\bea
\label{lgi1}
   \mathcal{L}_1(t) & = &  2P_{e \rightarrow e}(t) +2 P_{e \rightarrow e}(2t)-2 P_{e \rightarrow \mu}(t) \geq 0 \, , \\[2mm]
\label{lgi2}
   \mathcal{L}_2(t) & = &  2P_{e \rightarrow \mu}(t) - P_{e \rightarrow \mu}(2t) \geq 0 \, , \\[2mm]
   \label{lgi3}
   \mathcal{L}_3(t) & = &  2 P_{e \rightarrow e}(2t) \geq 0 \, , \\[2mm]
   \label{lgi4}
      \mathcal{L}_4(t) & = &  2P_{e \rightarrow \mu}(t) +2 P_{e \rightarrow \mu}(2t)-2 P_{e \rightarrow e}(t) \geq 0 \, . \eea
It is evident that Eq. \eqref{lgi3} is trivially satisfied. However, it is interesting to compare the other Leggett--Garg conditions with the NSIT~\eqref{NSIT:neutrino}. A plot of the above functions together with $\mathcal{N}(t)$ for the flavor oscillation probability~\eqref{OscWP} in the wave-packet description is displayed in Fig.~\ref{fig:my_nsitvslgi}. It can be immediately observed that the entire set of LGIs \eqref{lgi1}-\eqref{lgi4} is always fulfilled at late times, whilst $\mathcal{N}(t) \neq 0$. This divergence in the predictions could have been foreseen, since the NSIT \eqref{NSIT:neutrino} is a necessary and sufficient condition for macrorealism, but the LGIs \eqref{lgi1}-\eqref{lgi4} are not. 
%Moreover, for large $t$, when $\mathcal{N}(t) \to 0$, and macrorealistic picture can be approximately regarded as valid, all $\mathcal{L}_j$ are non-negative, confirming hence the previous results of Ref.~\cite{Shafaq:2021lju}, where only $\mathcal{L}_2$ has been studied in presence of different sources of decoherence. 
%

\begin{figure}[t]
\centering
\hspace{-5mm} 
    \includegraphics[width=9cm]{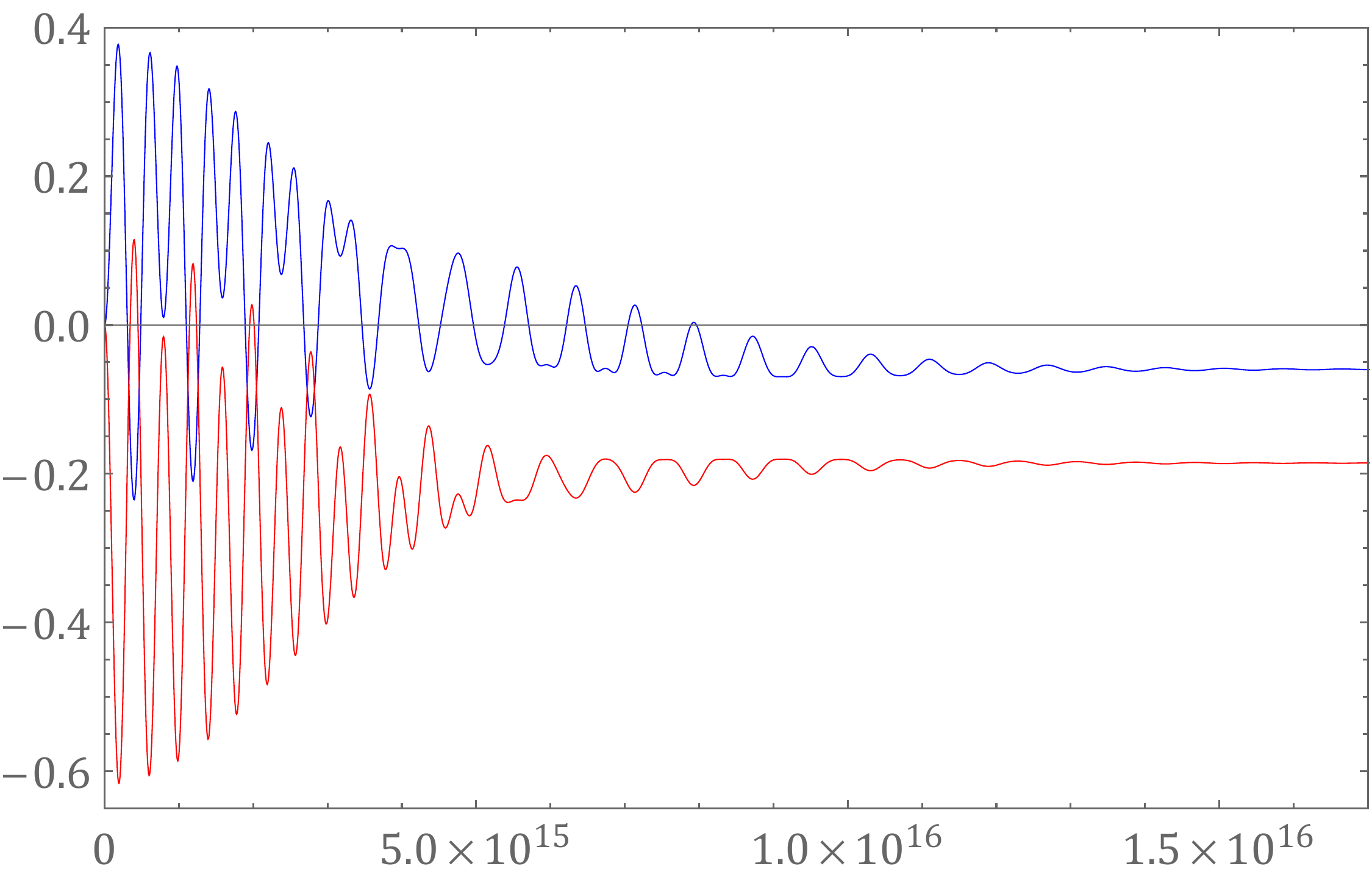}
    
\hspace{-5mm} 
    \includegraphics[width=9cm]{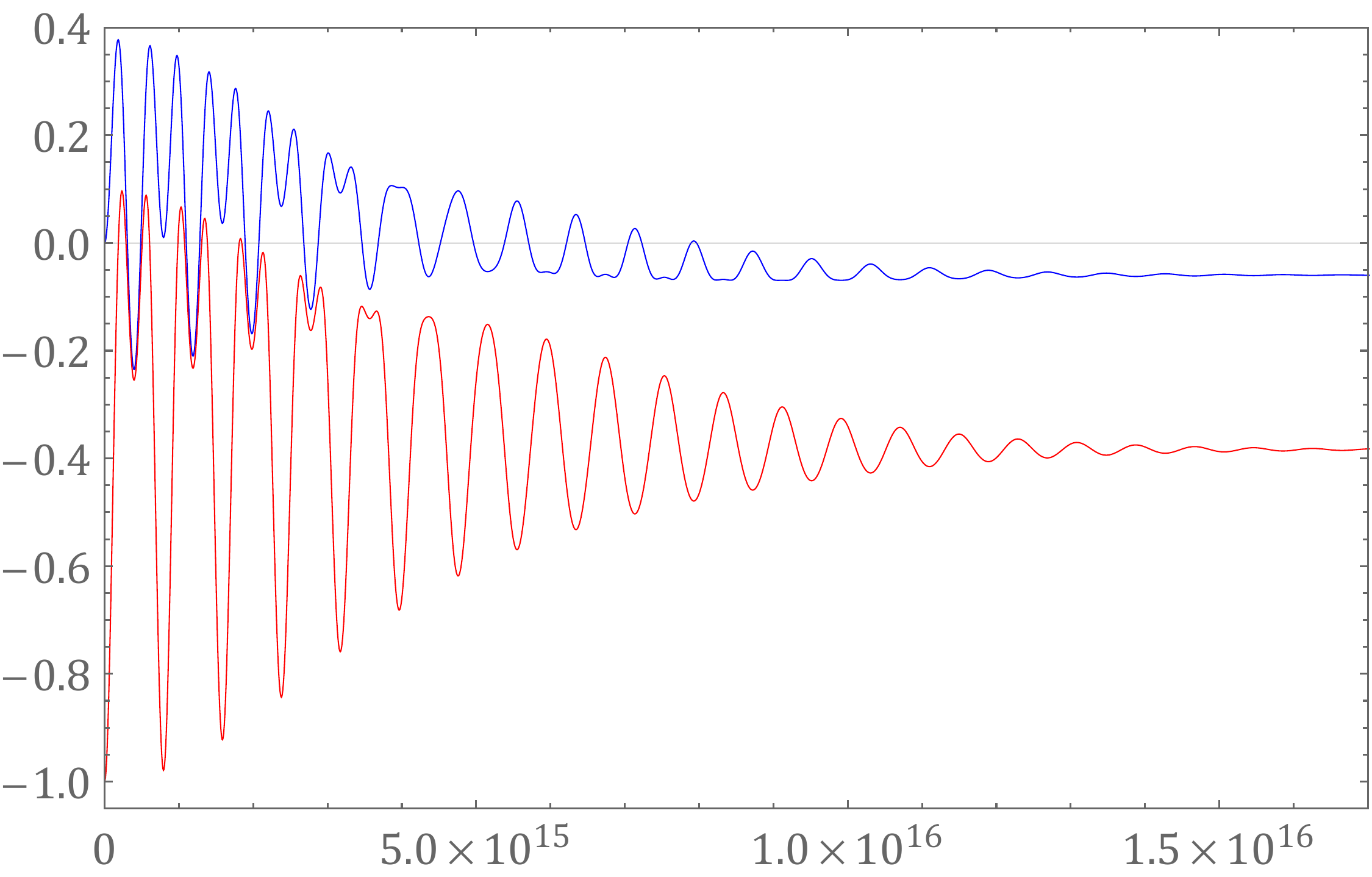}
    \caption{$\mathcal{N}(t)$ (blue) vs $\mathcal{W}_1(t)$ (red, first plot) and $\mathcal{W}_2(t)$ (red, second plot) as functions of time expressed in eV$^{-1}$. The former witnesses violation of the NSIT condition whenever it is not equal to zero, while the latter witness violation of WLGIs whenever they are positive. Note that for large $t$, all $\mathcal{W}_j(t)$ are always non-positive while $\mathcal{N}(t)$ differs from zero. The values used to generate the plot have been taken from the MINOS experiment \cite{data}, with $\sin^2\theta=0.314$, $\Delta m^2=7.92\times10^{-5}$ eV$^2$, $E=10$ GeV and $\sigma_x=0.5$ GeV$^{-1}$.}
    \label{fig:my_nsitvsWlgi}
\end{figure}

Turning the attention on the Wigner formulation of LGIs \eqref{wlgi}-\eqref{wlgi3}, we observe that they are already cast in terms of probabilities of measurement outcomes, and hence the identification with flavor oscillation probabilities turns out to be more natural. Indeed, we obtain
\begin{eqnarray}
\label{lgwn} \mathcal{W}_1(t) & = & P_{e \rightarrow e} (t) \, P_{\mu \rightarrow e} (t)-P_{\mu \rightarrow e} (2t) \ \leq \ 0 \, , \\
\label{lgwn2} \mathcal{W}_2(t) & = & P^2_{e \rightarrow \mu} (t)-P_{e \rightarrow e} (2t) \leq \ 0 \, , \\
\label{lgwn3} \mathcal{W}_3(t) & = & P_{e \rightarrow e} (t) \, P_{\mu \rightarrow e} (t)-P_{\mu \rightarrow e} (2t) \ \leq \ 0 \, ,
\end{eqnarray}
%\bea \label{lgwn}
%\hspace{-0.3cm} \mathcal{W}_1(t) \! & = & \! \mathcal{W}_3(t) \! = \!
%P_{e \rightarrow e} (t) \, P_{\mu \rightarrow e} (t)-P_{\mu \rightarrow e} (2t) \ \leq \ 0 \, , \\[2mm] \label{lgwn2}
%\hspace{-0.3cm} \mathcal{W}_2(t) \! & = & \! 
% P^2_{e \rightarrow \mu} (t)-P_{e \rightarrow e} (2t) \leq \ 0 \, .
%\eea
%
from which we deduce that the first and the last conditions coincide, thus leading to two non-trivial inequalities for neutrino oscillations. In Fig.~\ref{fig:my_nsitvsWlgi}, we compare the relevant WLGIs with the NSIT condition~\eqref{NSIT:neutrino}.
As in the case of the standard LGIs~\eqref{lgi1}--\eqref{lgi4}, the WLGIs~\eqref{lgwn}--\eqref{lgwn3} are satisfied for large $t$, where according to the NSIT condition no macrorealistic interpretation can be alleged, thereby confirming the previous considerations. In fact, the obtained results for both formulations of LGIs reveal that a macrorealistic description is not necessarily valid in the regime where such inequalities are satisfied.
%%%%%%%%%%%%%%%%%%%%%%%%%%%%%%%%%%%%%%%%%%%%%%%%%%%%%%%%%%%%%%%%%%%%%%%%%%%%%%%%%

\section{Conclusions}\label{sec:5}
In this paper, we have provided a preliminary analysis of necessary and sufficient conditions for macrorealism in neutrino flavor transitions. In particular, we have unambiguously found that the set of necessary and sufficient NSIT/AoT conditions derived in Ref.~\cite{Clemente2015} reduces to a single, non-trivial NSIT relation for macrorealism which can be potentially probed in two-flavor neutrino experiments. Moreover, concerning the wave-packet approach, we have seen that the effect of decoherence for long detection times/distances allows for a net deviation from a macrorealistic interpretation, thereby unambiguously attributing a quantum nature to the phenomenon of neutrino oscillations. For this reason, neutrinos can never be described in a macrorealistic way, even when quantum coherence is apparently degraded because of the wave packet spreading. 

Additionally, we have compared the aforementioned NSIT condition for macrorealism with the LGIs in their standard and Wigner formulations. In both scenarios, we have discovered that, as long as the NSIT requirement is met, the LGIs are satisfied.
However, at late times, we have shown that the LGIs are not faithful quantifiers of the macrorealistic description, since they are fulfilled whilst the NSIT condition is always violated.

Our research paves the way toward a more accurate study of macrorealism for neutrino flavor transitions. Although the phenomenology of neutrino oscillations can be effectively studied in the framework of quantum mechanics (QM), a proper treatment of neutrinos demands the application of quantum field theory (QFT) due to their relativistic nature~ \cite{Smaldone:2021mii}. As a preliminary analysis along this direction, in Ref.~\cite{Blasone:2021mbc} violations of the WLGIs in neutrino oscillations have been compared in the context of QM and QFT. Interestingly, it turns out that QFT violates the WLGIs more frequently than QM, which is in agreement with the results obtained for the Bell tests of local realism within the general framework of algebraic QFT~\cite{Summers:1987fn,Summers:1987fepr,Summers:1987ze}. As a further evidence for the same occurrence, it has been proven that even vacuum correlations in QFT can lead to a maximal quantum violation of Bell inequalities \cite{ill6}. Therefore, both studies seem to indicate that QFT is \emph{less classical} than QM, and these results has to be reviewed by means of the NSIT/AoT conditions for macrorealism. 
%%%%%%%%%%%%%%%%%%%%%%%%%%%%%%%%%%%%%%%%%%%%%%%%%%%%%%%%%%%%%%%%%%%%
\section*{Acknowledgements}
F.I. and L.P. acknowledge support by MUR (Ministero dell'Universit\`a e della Ricerca) via the project PRIN 2017 ``Taming complexity via QUantum Strategies: a Hybrid Integrated Photonic approach'' (QUSHIP) Id. 2017SRNBRK. L.S. acknowledges support by the Polish National Science Center grant 2018/31/D/ST2/02048. L.P. acknowledges networking support by the COST Action CA18108 and is grateful to the ``Angelo Della Riccia'' foundation for the awarded fellowship received to support the study at Universit\"at Ulm.

%\section*{References}
%%%%%%%%%%%%%%%%%%%%%%%%%%%%%%%%%%%%%%%%%%%%%%%%%%%%%

\bibliography{LibraryNeutrino}

\bibliographystyle{apsrev4-2}

\end{document}